\documentclass[journal]{IEEEtran}
\usepackage{url}
\usepackage[hidelinks]{hyperref}

\usepackage{cite}
\usepackage{amsmath,amssymb,amsfonts}
\usepackage{graphicx}
\usepackage{textcomp}
\usepackage{float}
\usepackage{enumitem}
\usepackage{stfloats}
\usepackage{bm}

\def\BibTeX{{\rm B\kern-.05em{\sc i\kern-.025em b}\kern-.08em
    T\kern-.1667em\lower.7ex\hbox{E}\kern-.125emX}}

\graphicspath{{./media/}}

\begin{document}

\title{Multimodal User Authentication via Keystroke-Triggered Fusion of Glove-Based Finger Pressure and Hand Kinematics}

\author{Issei~Hyakuda and Lei~Jing,~\IEEEmembership{Member,~IEEE}%
\thanks{This article has been published in \emph{IEEE Access}, vol.~14, pp.~163209--163221, 2026, DOI: \href{https://doi.org/10.1109/ACCESS.2026.3736691}{10.1109/ACCESS.2026.3736691}.}%
\thanks{The authors are with the Graduate School of Computer Science and Engineering, The University of Aizu, Aizuwakamatsu 965-8580, Japan (e-mail: m5301024@u-aizu.ac.jp; leijing@u-aizu.ac.jp).}%
\thanks{Corresponding author: Lei Jing (e-mail: leijing@u-aizu.ac.jp).}}

\markboth{IEEE Access, vol.~14, 2026}%
{Hyakuda and Jing: Multimodal User Authentication via Keystroke-Triggered Fusion of Glove-Based Finger Pressure and Hand Kinematics}

\maketitle

\begin{abstract}
Keystroke dynamics constitute a cost-effective behavioral biometric, but their timing features degrade under environmental and hardware variation. We replace these features with 19 dimensions of finger pressure and hand kinematics, sampled around each keystroke by a custom data glove -- ten piezoresistive pressure sensors and a nine-axis inertial measurement unit (IMU). The two modalities are encoded by separate convolutional neural network--long short-term memory (CNN-LSTM) branches and joined only at the embedding. We collected typing data from ten participants, one enrolled as the genuine user. Training uses a desktop keyboard and evaluation a laptop keyboard recorded on a later day; the held-out impostor is excluded from training, and each of the nine takes that role in turn (Leave-One-Impostor-Out). Under this condition the method attains a mean Equal Error Rate (EER) of 0.53 $\pm$ 0.42\% after three consecutive keystrokes, about 0.77 s of typing at the measured median pace, and 0.18 $\pm$ 0.22\% after five. Results depend strongly on which impostor is held out. A per-modality ablation finds the fused model significantly better than both single-modality models at every smoothing window. The pressure-level channels carry that advantage and have to be scale-matched to deliver it, by separate encoders or by standardization indifferently.
\end{abstract}

\begin{IEEEkeywords}
Behavioral biometrics, biometric authentication, convolutional neural networks, data fusion, data glove, deep learning, keystroke dynamics, long short-term memory (LSTM), multimodal authentication, wearable sensors.
\end{IEEEkeywords}

\section*{Nomenclature}
\begin{description}[labelwidth=1.8cm, leftmargin=2.0cm, style=multiline,
                    font=\normalfont, itemsep=1pt, parsep=0pt, topsep=3pt]
\item[$A_x, A_y, A_z$] Triaxial acceleration from the glove IMU (units: $g$).
\item[$C$] Number of channels supplied to the classifier ($C=29$: 19 normalized sensor channels and 10 pressure-level channels).
\item[$C_{p}$] Number of appended pressure-level channels ($C_{p}=10$).
\item[$d$] LSTM hidden size ($d=64$ units).
\item[EER] Equal Error Rate (\%), the operating point at which FAR = FRR.
\item[FAR] False Acceptance Rate (\%).
\item[FRR] False Rejection Rate (\%).
\item[$G_x, G_y, G_z$] Triaxial angular velocity from the glove IMU (units: $^{\circ}/\text{s}$).
\item[$L$] Fixed sequence length after linear interpolation ($L=45$ time steps).
\item[$\ell_{c}$] Pressure level of pressure channel $c$ within a window, $\mu_{c}/\sigma_{c}$; constant along the time axis.
\item[LOIO] Leave-One-Impostor-Out, the evaluation protocol of Section \ref{sec:data_collection}: the enrolled genuine user is fixed and the held-out impostor is rotated over the remaining participants.
\item[$M_x, M_y, M_z$] Triaxial geomagnetism from the glove IMU (units: $\mu$T).
\item[$p$] Dropout rate ($p=0.395$).
\item[$s$] Per-keystroke authentication score, the softmax probability of the genuine class.
\item[$\bar{s}_{t}$] Smoothed score at keystroke $t$: the trailing mean of $s$ over $W$ keystrokes.
\item[$T_{d}$] End-to-end latency of a smoothed authentication decision (units: s).
\item[$T_{w}$] Per-keystroke feature-extraction window length (units: ms; $T_{w}=600$\,ms unless stated otherwise).
\item[$W$] Moving-average smoothing window size (number of consecutive keystrokes); distinct from $T_{w}$.
\item[$\hat{\mathbf{X}}$] Normalized feature window supplied to the classifier ($L \times C$); $\hat{\mathbf{X}}_{p}$ and $\hat{\mathbf{X}}_{m}$ are its pressure and inertial parts.
\item[$y$] Class label; $y=0$ denotes an impostor and $y=1$ a genuine window.
\item[$\Delta t_{k}$] Inter-keystroke interval between consecutive key presses (units: ms).
\item[$\kappa$] Convolution kernel width ($\kappa = 3$).
\item[$\mu, \sigma$] Mean and standard deviation used for per-window Z-score normalization.
\end{description}

\section{Introduction}
\label{sec:introduction}
\IEEEPARstart{C}{lassical} biometrics such as fingerprint recognition achieve low error rates at the point of login, but they are poorly suited to continuous background authentication. 

Keystroke Dynamics (KD) offers a cost-effective and passive means of continuous authentication, but unimodal KD relies exclusively on temporal features -- dwell and flight times -- which renders it susceptible to environmental fluctuation, user fatigue and presentation attacks such as robotic spoofing. Overcoming these limitations calls for physical traits beyond timing, yet the synchronous acquisition of such heterogeneous signal streams remains challenging.

We therefore sense at the hand rather than at the keyboard, and supply no timing features to the classifier. An attack that reproduces a victim's dwell and flight times \cite{Yu2025} accordingly conveys nothing to it. We have not, however, evaluated attacks that attempt to reproduce the contact pressure and hand kinematics themselves, and the signals are not thereby invariant to the keyboard: cross-device generalization is something the model must still learn, and testing it is one purpose of the evaluation in Section \ref{sec:experimental_results}.

Sensing fine-grained finger pressure and 3D hand kinematics with a glove has not been evaluated for continuous authentication (Section \ref{sec:related_wearable}). This raises our research question: can these signals, captured at each keystroke in place of timing features, restore robustness under the combined conditions of environmental and cross-device variation and unseen impostors, under which unimodal KD fails?

To answer it we propose a multimodal framework that treats keystroke events as temporal anchors, classifies the 19-dimensional pressure and kinematic signals captured in their vicinity, and aggregates the per-keystroke scores over consecutive keystrokes. The evaluation is offline.

To obtain these signals synchronously and at high resolution we constructed a custom data glove. It is a proof-of-concept platform rather than a consumer device, although the form factor is already practical in VR/AR controllers and in high-security industrial and access-control settings. A wrist-worn IMU alone would not reach per-finger pressure, so the pressure sensing must either stay on the hand -- as a partial glove or a ring-shaped device -- or move to a pressure-sensitive keyboard, which is the direction miniaturization would have to take.

The main contributions of this paper are summarized as follows:
\begin{itemize}
    \item \textbf{Multimodal wearable integration:} We designed a custom data glove with ten piezoresistive pressure sensors and a nine-axis IMU.
    \item \textbf{Timing-free multimodal fusion:} We fuse contact pressure and 3D hand motion at each keystroke and supply no dwell or flight times to the classifier.
    \item \textbf{Pressure-level scale matching:} Matching the pressure-level channels to the scale of the others is what makes them usable, and encoding the modalities separately or standardizing the block does it equally well.
    \item \textbf{Unseen-impostor, cross-domain evaluation:} We evaluate from desktop to laptop with the held-out impostor rotated over all nine remaining participants, obtaining a mean Equal Error Rate (EER) of 0.53 $\pm$ 0.42\% after roughly 0.77 seconds of typing.
\end{itemize}

The remainder of this paper is organized as follows. Section \ref{sec:related_work} reviews related work. Section \ref{sec:proposed_method} details the proposed multimodal method and the supporting hardware. Section \ref{sec:evaluation} presents the data collection setup, the ablation studies, and the empirical results. Section \ref{sec:discussion} discusses the implications of these results. Finally, Section \ref{sec:conclusion} concludes the paper.

\section{Related Work}
\label{sec:related_work}

\subsection{Authentication Based on Keystroke Dynamics (Unimodal KD)}
\begin{figure}[htbp]
\centering
\includegraphics[width=1.0\linewidth]{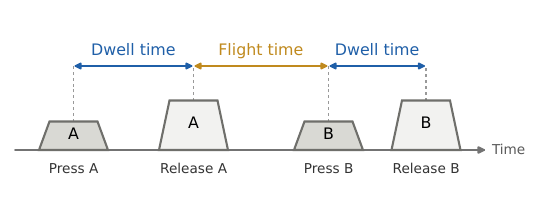}
\caption{Temporal features in Keystroke Dynamics: Dwell Time and Flight Time}
\label{fig:2.1}
\end{figure}

Keystroke Dynamics (KD) authenticates users on the basis of their typing rhythms. As established in foundational studies \cite{Roy2022}, classical KD features consist principally of dwell time (the duration of a key press) and flight time (the interval between releasing one key and pressing the next), as illustrated in Figure \ref{fig:2.1}. 

Many machine learning and deep learning models have been applied to these temporal features, but performance still degrades with environmental variation, keyboard hardware and user fatigue \cite{Shadman2025}. Yu et al.\ \cite{Yu2025} further showed unimodal KD to be vulnerable to ``robot adversarial attacks,'' in which automated agents use generative models to mimic human typing rhythms.

\subsection{Device-Centric Physical Sensing Approaches}\label{sec:related_device_centric}
To address the vulnerabilities of timing-only KD, researchers have incorporated physical sensing through device-centric approaches, principally by means of mobile devices and instrumented keyboards. Smartphone-centric methods exploit built-in sensors to capture interaction dynamics \cite{rayani2023continuous}. The pioneering HMOG framework \cite{sitova2016hmog}, for example, fused keystroke data with spatial sensor data and thereby lowered authentication errors substantially, and a more recent study has analyzed touchscreen swipe biometrics \cite{al2023keystroke}.
These mobile approaches, however, capture only macro-level device motion or two-dimensional surface interaction, and therefore miss the fine-grained finger articulation of the user.

As an alternative, customized instrumented keyboards and smart touch interfaces have been proposed in order to capture typing force and spatial interaction directly. Chen et al.\ \cite{chen2015personalized} developed an intelligent keyboard that exploits contact electrification to trace personalized keystroke force, whereas Zhang et al.\ \cite{Zhang2025} integrated flexible pressure-sensor arrays into the typing interface. The multifunctional touch sensor of Wei et al.\ \cite{wei2022all} combines carbon-based gradient resistance elements with deep learning for biometric verification. These interfaces capture high-resolution physical traits, but require modified input hardware, which restricts cross-device use -- moving from a desktop keyboard to a laptop, for instance.

\subsection{Wearable Hand Motion Biometrics}\label{sec:related_wearable}
Before wearable devices became widespread, noncontact methods of capturing hand kinematics were explored. Vision-based approaches \cite{roth2014continuous} track hand movement during typing by means of a webcam. Although such systems avoid hardware modification, they are susceptible to visual occlusion and to changes in illumination, raise the privacy concerns associated with continuous video recording, and cannot capture the physical contact pressure of keystrokes. 

Wearable sensors offer a direct and privacy-preserving alternative. Wrist-worn devices have been used for physiological biometrics such as cardiac signals \cite{Zhao2022}, and wearable sensors are under active study for hand motion and gesture recognition \cite{Pan2023}, \cite{Pyun2024}. In the specific domain of continuous typing authentication, Acar et al.\ \cite{acar2020usable} proposed a wearable-assisted framework based on the built-in IMU of a smartwatch. Although this work demonstrates the viability of wearable IMUs, wrist-worn devices typically capture macro-level arm motion and thus fail to detect the independent microkinematics of individual fingers. 

Closer to keystroke authentication, Senarath et al.\ proposed BehaveFormer \cite{Senarath2023}, a spatio-temporal dual-attention Transformer fusing smartphone IMU signals with keystroke and touch dynamics. It demonstrates the value of combining attention-based temporal modeling with IMU signals, but relies on a smartphone's built-in IMU and targets a touchscreen rather than a physical keyboard. Section \ref{sec:comparative_ablation} compares the proposed method against an adaptation of the authors' reference implementation under our own protocol.

Highly articulated wearables such as multisensor data gloves have likewise been used for behavioral biometrics, dynamic signature verification \cite{sayeed2006dynamic} being the best-known example. To the best of our knowledge, however, no study has captured finger pressure and 3D hand kinematics together with a glove during typing, fused them, and evaluated the result in a continuous-authentication setting. That is the setting this paper addresses.


\section{Proposed Method}
\label{sec:proposed_method}
\subsection{System Overview}
\begin{figure}[htbp]
\centering
\includegraphics[width=0.77\linewidth]{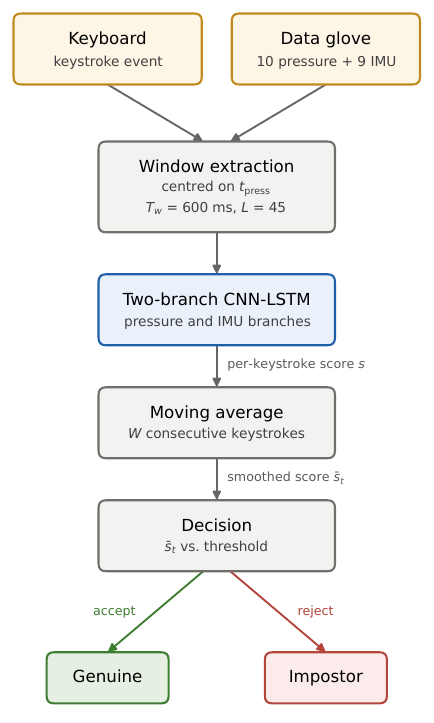}
\caption{The proposed authentication pipeline.}
\label{fig:3.1}
\end{figure}

Figure \ref{fig:3.1} shows the pipeline. The custom data glove is the only hardware the method adds; every stage after it is software.

The workflow is as follows:
\begin{enumerate}
    \item \textbf{Sensing:} The user wears the glove on the right hand while typing, and the device captures contact pressure and hand motion (Section \ref{sec:hardware}). The keyboard contributes the key-press timestamps and nothing else.
    \item \textbf{Window extraction:} Raw sensor data are filtered, normalized and synchronized with keystroke events by timestamp, then segmented into windows of length $T_{w}$ centered on each key press and interpolated to a common length $L$ (Section \ref{sec:preprocessing}).
    \item \textbf{Two-branch CNN-LSTM:} The pressure and inertial channels are encoded separately and joined only at the embedding, yielding a per-keystroke score $s$ (Section \ref{sec:model}).
    \item \textbf{Moving average and decision:} Consecutive scores are averaged over $W$ keystrokes (Section \ref{sec:postprocessing}) and the smoothed score $\bar{s}_{t}$ is compared with a threshold to accept or reject.
\end{enumerate}

\subsection{Sensing: The Data Glove}\label{sec:hardware}
To capture physical traits that standard keyboards cannot measure, we developed a sensor-equipped data glove.

\subsubsection{Sensors Configuration}
\begin{figure}[htbp]
\centering
\includegraphics[width=0.8\linewidth]{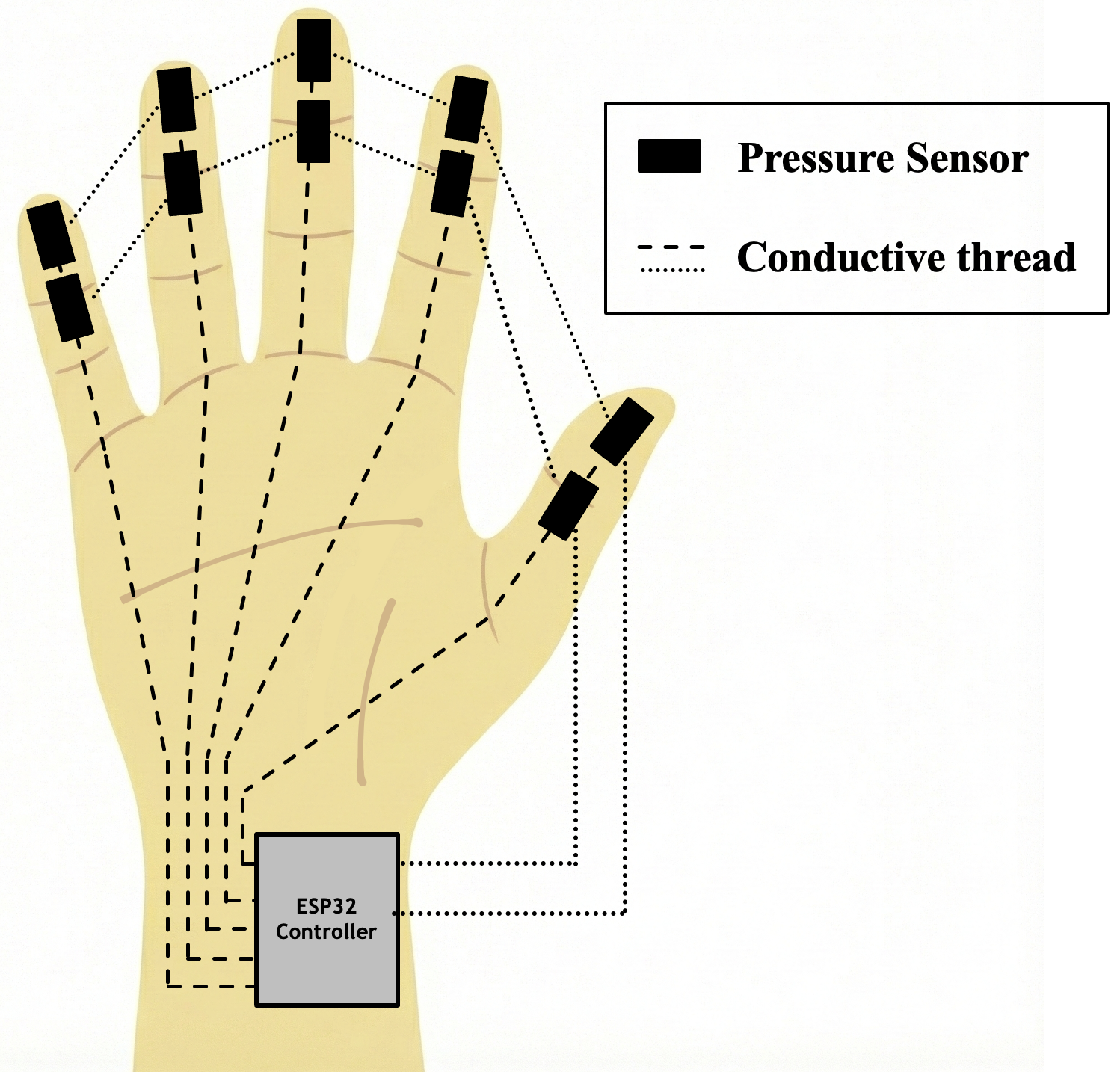}
\caption{Sensor placement on the data glove}
\label{fig:3.2.1}
\end{figure}

The glove was designed for the right hand and equipped with two types of sensor in order to acquire a 19-dimensional feature vector. The sensor placement is illustrated in Figure \ref{fig:3.2.1}.

\begin{itemize}
    \item \textbf{Pressure sensors (10 channels):} Each sensor is a piece of Velostat, a piezoresistive conductive material, sandwiched between conductive threads stitched vertically and horizontally into the glove. Ten are distributed over the hand, five at the fingertips and five at the distal interphalangeal joints, capturing changes in the pressure distribution during key presses and finger flexion.
    
    \item \textbf{Control unit and IMU (9 channels):} A custom ESP32-based controller on the palmar side of the wrist is the hub for data acquisition. Besides collecting and transmitting the pressure data it integrates a nine-axis IMU recording triaxial acceleration ($A_x, A_y, A_z$), angular velocity ($G_x, G_y, G_z$) and geomagnetism ($M_x, M_y, M_z$) throughout the intervals between key presses.
\end{itemize}

\subsubsection{Sampling, Synchronization, and Latency}
\label{sec:sync}
A single onboard ESP32 samples the ten pressure channels and the nine-axis IMU on one hardware timer interrupt at 100\,Hz and packs them into one UDP packet, so the two modalities share a sampling clock. The glove does not timestamp keystrokes; the host PC clock timestamps both the key events and the arrival of each packet, so no cross-device drift arises either.

Measured over the corpus, the inter-packet interval averages $10.12$\,ms, matching the nominal rate, with a standard deviation of $12.05$\,ms. That spread is host-side arrival jitter -- transport and operating system delivery -- and does not affect the samples within a packet, which carry the shared onboard clock. It is smaller than one step of the interpolated time axis the classifier sees ($600/45 = 13.3$\,ms), so on that axis it amounts to a displacement of less than a single step.

\subsubsection{Circuit Design}
\begin{figure}[htbp]
\centering
\includegraphics[width=0.7\linewidth]{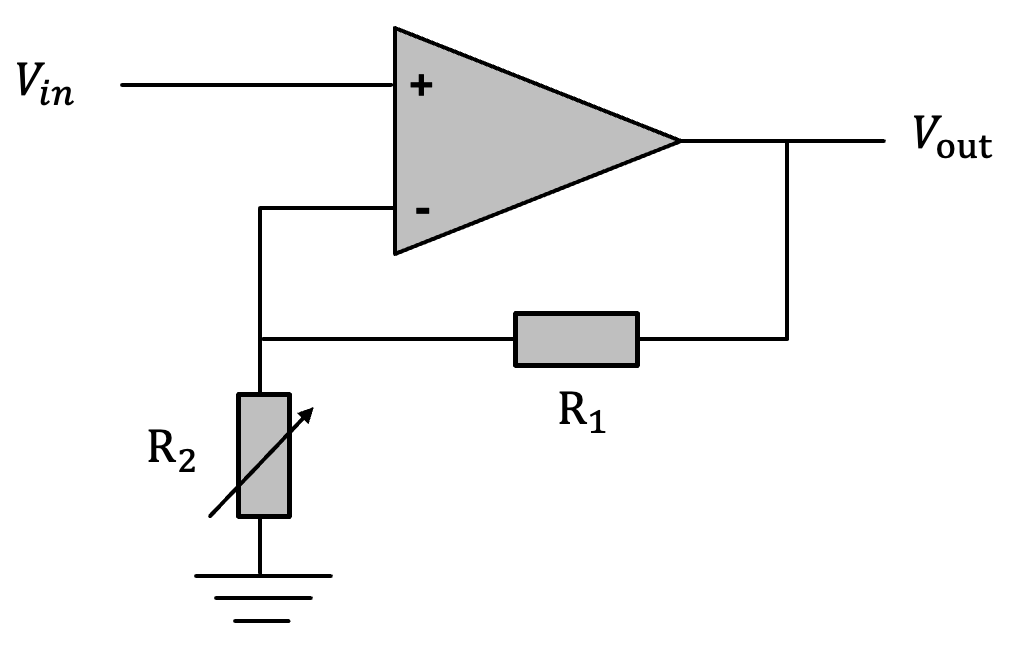}
\caption{Noninverting amplifier circuit for pressure sensors}
\label{fig:3.2.2}
\end{figure}

In order to read the piezoresistive Velostat sensors with high precision, we implemented the noninverting operational-amplifier circuit shown in Figure \ref{fig:3.2.2}.

In this circuit, the output voltage $V_{\mathrm{out}}$ produced by an input signal voltage $V_{\mathrm{in}}$ is determined by the ratio of the feedback resistor $R_1$ to the ground resistor $R_2$, as expressed in \eqref{eq:1}:

\begin{equation}
V_{\mathrm{out}} = \left( 1 + \frac{R_1}{R_2} \right)V_{\mathrm{in}}. \label{eq:1}
\end{equation}

This configuration presents a high input impedance and amplifies the minute resistance changes induced in the Velostat by typing pressure into a voltage range detectable by the analog-to-digital converter of the ESP32.

\subsection{Window Extraction and Preprocessing}\label{sec:preprocessing}
To address noise and variable sequence lengths in the raw sensor data, we extract a window around each key press and apply the following preprocessing steps, all keyed to the keystroke timing.

\subsubsection{Noise Reduction}
The pressure sensors occasionally produce nonphysical readings, such as saturation or open-circuit artifacts. Their normal operating range is around 300 ADC counts whereas these artifacts reach several times that, so prior to windowing any reading of $1000$ counts or more was clipped to the largest valid value observed in the corresponding session. The rule affects $0.03\%$ of all pressure samples.

\subsubsection{Window Extraction and Interpolation}
\label{sec:segmentation}
Because typing speed varies, keystroke events differ in duration. We therefore extract a fixed-duration window centered on the key-press timestamp $t_{\mathrm{press}}$, where $T_{w}=600$\,ms was selected by the Optuna search described in Section \ref{sec:optimization}.

All glove samples whose timestamps fall inside this interval form the raw segment for that keystroke, in their original order; a segment of fewer than five samples is discarded as too sparse to interpolate. The segment is then resampled to $L=45$ steps by piecewise linear interpolation over a normalized time axis, independently per channel, giving the CNN a fixed input dimension of $L \times C$.

Keystroke timestamps serve only to place the window: neither dwell nor flight times are supplied to the classifier, whose input comes entirely from the glove channels. Section \ref{sec:normalization} appends $C_{p}=10$ channels computed from the same pressure signals, bringing the input to $C=29$.

Adjacent windows overlap whenever consecutive keystrokes fall closer together than $T_{w}$; Section \ref{sec:window_analysis} quantifies this overlap and reports a sensitivity analysis with respect to $T_{w}$.

\subsubsection{Normalization}
\label{sec:normalization}
The value ranges of the pressure (resistance) channels and the IMU (acceleration and angular velocity) channels differ by orders of magnitude. To eliminate this scale bias, we applied Z-score normalization, as given in \eqref{eq:norm}:
\begin{equation}
x_{\mathrm{norm}} = \frac{x - \mu}{\sigma}, \label{eq:norm}
\end{equation}
where $\mu$ and $\sigma$ are the mean and standard deviation of one channel within one feature window, computed along the time axis of that window alone. No statistic is shared across windows, across channels, or between the training and test partitions, so the normalization introduces no dependence between samples. In the implementation $\sigma$ is floored at $10^{-6}$, which guards against a channel that stays constant throughout a window, as happens for fingers not in contact with the keyboard.

Subtracting $\mu$ removes the absolute level of every channel. The inertial channels lose little by this, but the pressure channels lose how firmly each finger rests and strikes. Retaining that level instead of centering it improves a pressure-only model from 13.49\% to 2.69\% EER at $W=5$. We therefore restore the quantity explicitly, as $C_{p}=10$ additional channels, which reaches 3.16\% under the same protocol (Table \ref{tab:modality_ablation}) while leaving the per-window Z-score of the other channels intact:
\begin{equation}
\ell_{c} = \frac{\mu_{c}}{\sigma_{c}}, \qquad c \in \{1,\dots,10\}, \label{eq:level}
\end{equation}
one per pressure channel, held constant along the time axis and appended to the 19 normalized channels, so that the classifier receives $C=29$ in total. The level is expressed as a ratio to the within-window variation rather than as a raw mean because a change of glove fit or amplifier gain scales $\mu$ and $\sigma$ together and leaves the ratio comparatively fixed. Like $\mu$ and $\sigma$, $\ell$ is computed from a single window.

\subsection{Two-Branch CNN-LSTM Model}\label{sec:model}
\begin{figure}[htbp]
    \centering
    \includegraphics[width=0.81\linewidth]{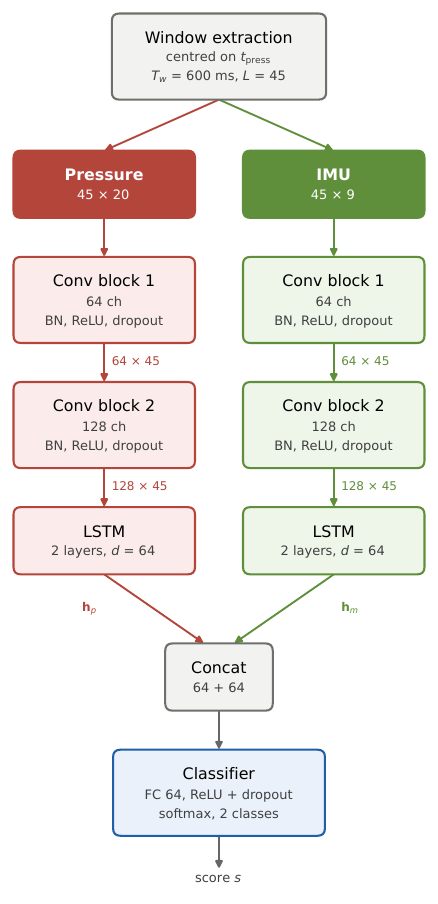}
    \caption{Architecture of the proposed keystroke-triggered two-branch CNN-LSTM model.}
    \label{fig:3.4}
\end{figure}

To learn the correlation between physical pressure and hand motion, we developed a hybrid deep learning model that combines a Convolutional Neural Network (CNN) \cite{Lecun2002} with a Long Short-Term Memory (LSTM) network \cite{Hochreiter1997}. The detailed architecture is shown in Figure \ref{fig:3.4}.

The model encodes the two modalities separately and joins them only at the embedding. Two identical branches, one taking the $20$ pressure channels of Section \ref{sec:normalization} and the other the $9$ inertial channels, each apply the CNN and LSTM stages described below; their final hidden states are concatenated and passed to a single classification head.

The arrangement is not incidental. A single stream over the concatenated channels -- the obvious design -- lets one modality dominate the filters the first convolution learns, because that layer mixes all channels before batch normalization sees them. Encoding separately removes the interaction: nothing is mixed until both branches have produced a fixed-length embedding. Sections \ref{sec:comparative_ablation} and \ref{sec:modality_ablation} report what the separation is worth.

The feature-window length, the LSTM hidden size and the learning rate were selected with the Optuna framework, as described in Section \ref{sec:optimization}; the convolution kernel width was fixed at $\kappa=3$, the LSTM depth at two layers, and the dropout rate at $p=0.395$.

Write $\hat{\mathbf{X}} \in \mathbb{R}^{L \times C}$ for the normalized window, with $L=45$ and $C=29$, and let $\hat{\mathbf{X}}_{p} \in \mathbb{R}^{L \times 20}$ and $\hat{\mathbf{X}}_{m} \in \mathbb{R}^{L \times 9}$ be its pressure and inertial parts. Each branch $\beta \in \{p, m\}$ processes its own part through the three stages below, producing a read-out vector $\mathbf{h}_{\beta} \in \mathbb{R}^{d}$; the stages are described once, since the two branches differ only in their input width and share no weights.

\subsubsection{Feature Extraction (CNN Block 1 \& 2)}
The first stage consists of two one-dimensional (1-D) convolutional blocks designed to extract spatial, cross-channel features from the sensor signals while preserving the temporal dimension. Within a branch the input is transposed to the channel-first layout $\mathbf{Z}^{(0)} = \hat{\mathbf{X}}_{\beta}^{\top} \in \mathbb{R}^{C_{0} \times L}$, with $C_{0}=20$ for the pressure branch and $C_{0}=9$ for the inertial branch.

Each block $b \in \{1,2\}$ applies a convolution, batch normalization, a rectified linear unit, and dropout, in that order:
\begin{equation}
\mathbf{Z}^{(b)} = \mathcal{D}_{p}\Bigl( \mathrm{ReLU}\bigl( \mathrm{BN}( \boldsymbol{\omega}^{(b)} * \mathbf{Z}^{(b-1)} + \boldsymbol{\beta}^{(b)} ) \bigr) \Bigr). \label{eq:cnnblock}
\end{equation}
Here $*$ denotes 1-D convolution along the time axis with kernel width $\kappa = 3$, unit stride, and zero padding of one step on each side, so that the temporal length $L$ is preserved; $\mathrm{BN}(\cdot)$ is batch normalization applied per output channel, with statistics taken over the minibatch and the time axis jointly; and $\mathcal{D}_{p}$ is dropout at rate $p = 0.395$, applied elementwise during training and replaced by the identity at inference. With $(C_{1},C_{2}) = (64,128)$ output channels, the two blocks map $C_{0} \times 45 \rightarrow 64 \times 45 \rightarrow 128 \times 45$, expanding the channel depth while leaving the temporal resolution untouched.

\subsubsection{Temporal Learning (LSTM Block)}
The feature map produced by the CNN blocks is transposed back to the time-first layout $\mathbf{H}^{(0)} = (\mathbf{Z}^{(2)})^{\top} \in \mathbb{R}^{L \times 128}$ and fed into a two-layer stacked unidirectional LSTM with hidden size $d = 64$, which models the temporal dependencies.

For layer $n \in \{1,2\}$ and time step $t = 1,\dots,L$, writing $\mathbf{h}^{(n-1)}_{t}$ for the input to that layer, the four gates are computed in a single affine map,
\begin{equation}
\begin{bmatrix} \mathbf{i}_{t} \\ \mathbf{f}_{t} \\ \mathbf{g}_{t} \\ \mathbf{o}_{t} \end{bmatrix}
= \begin{bmatrix} \varsigma \\ \varsigma \\ \tanh \\ \varsigma \end{bmatrix}
\Bigl( \mathbf{W}^{(n)}_{x}\mathbf{h}^{(n-1)}_{t} + \mathbf{W}^{(n)}_{h}\mathbf{h}^{(n)}_{t-1} + \mathbf{b}^{(n)} \Bigr), \label{eq:lstm_gates}
\end{equation}
and the cell and hidden states are updated as
\begin{equation}
\mathbf{c}^{(n)}_{t} = \mathbf{f}_{t} \odot \mathbf{c}^{(n)}_{t-1} + \mathbf{i}_{t} \odot \mathbf{g}_{t}, \qquad
\mathbf{h}^{(n)}_{t} = \mathbf{o}_{t} \odot \tanh \mathbf{c}^{(n)}_{t}, \label{eq:lstm_state}
\end{equation}
where $\varsigma(\cdot)$ is the logistic sigmoid applied rowwise, $\odot$ denotes the Hadamard product, and $\mathbf{h}^{(n)}_{0} = \mathbf{c}^{(n)}_{0} = \mathbf{0}$. The input to the first layer is $\mathbf{h}^{(0)}_{t} = \mathbf{H}^{(0)}[t,:]$; the input to the second is the dropout-perturbed output of the first, at the same rate $p = 0.395$, with no dropout applied to the output of the final layer.

The sequence is processed in a many-to-one fashion: only the final hidden state of the topmost layer, $\mathbf{h}_{\beta} = \mathbf{h}^{(2)}_{L} \in \mathbb{R}^{d}$, is retained, which condenses that branch's time series into a $64$-dimensional static feature vector.

\subsubsection{Fusion and Scoring}
The two branches meet here and nowhere earlier. Their read-out vectors are concatenated,
\begin{equation}
\mathbf{h} = \bigl[\, \mathbf{h}_{p} \;;\; \mathbf{h}_{m} \,\bigr] \in \mathbb{R}^{2d}, \label{eq:fusion}
\end{equation}
and the result is passed through a fully connected layer with ReLU activation and dropout, then through an output layer mapping to the two classes ``Impostor'' ($y=0$) and ``Genuine'' ($y=1$). The probability a softmax assigns to the genuine class is the per-keystroke authentication score:
\begin{equation}
\begin{aligned}
\mathbf{u} &= \mathcal{D}_{p}\bigl( \mathrm{ReLU}(\mathbf{W}_{1}\mathbf{h} + \mathbf{b}_{1}) \bigr), \\
s &= P(y=1 \mid \hat{\mathbf{X}}) = \mathrm{softmax}\bigl( \mathbf{W}_{2}\mathbf{u} + \mathbf{b}_{2} \bigr)_{1},
\end{aligned} \label{eq:softmax}
\end{equation}
with $\mathbf{W}_{1} \in \mathbb{R}^{64 \times 2d}$ and $\mathbf{W}_{2} \in \mathbb{R}^{2 \times 64}$. The whole network holds $230{,}146$ trainable parameters, fewer than the $303{,}298$ of a single-stream model with its hidden size doubled to $d=128$ to match the fused embedding.

\subsection{Implementation and Optimization}
\label{sec:optimization}

\subsubsection{Training Objective}
The network is trained by minimizing a class-weighted cross-entropy loss over a minibatch $\mathcal{B}$ with logits $\mathbf{z}_{n}$ and labels $y_{n} \in \{0,1\}$:
\begin{equation}
\mathcal{L}(\mathcal{B}) = \frac{\sum_{n \in \mathcal{B}} w_{y_{n}}\,\bigl[-\log \mathrm{softmax}(\mathbf{z}_{n})_{y_{n}}\bigr]}{\sum_{n \in \mathcal{B}} w_{y_{n}}}. \label{eq:loss}
\end{equation}
Because the impostor pool contributes many more windows than the single genuine user, the classes are unbalanced; the weights compensate for this in proportion to the observed counts, with $w_{0}=1$ for the impostor class and $w_{1} = N_{\mathrm{imp}}/N_{\mathrm{gen}}$ for the genuine class, where $N_{\mathrm{gen}}$ and $N_{\mathrm{imp}}$ are the numbers of genuine and impostor windows in the training partition. Optimization uses AdamW with a learning rate of $1.814 \times 10^{-3}$ and a batch size of $64$, for at most $100$ epochs. Training halts early when the validation loss has failed to improve for $8$ consecutive epochs, and the parameters restored for evaluation are those of the epoch attaining the lowest validation loss; the validation partition used for this purpose is defined in Section \ref{sec:data_collection}. Training additionally employs gradient clipping at a maximum norm of 5.0.

\subsubsection{Hyperparameter Search}
The configuration reported here was fixed by an initial search with Optuna \cite{Akiba2019} over the feature-window length $T_{w}$, the interpolated sequence length $L$, the LSTM hidden size $d$ and the learning rate, selecting $T_{w}=600$\,ms, $L=45$, $d=64$ and $1.814 \times 10^{-3}$. The layer count, dropout rate, weight decay and batch size were not searched: the two LSTM layers and $\kappa=3$ were fixed throughout, and $p=0.395$ was set separately and carried forward unchanged into every model reported here.

The validation loader of that search was built from the genuine cross-domain session on which this paper also reports, so its objective was not independent of the reported evaluation data. We repeated it under a partially nested protocol, with the reported session withheld. It chose a smaller model ($101{,}634$ parameters against $230{,}146$) that the nine folds cannot separate from the published one: 0.56\% against 0.53\% at $W=3$, Wilcoxon signed-rank $p \ge 0.20$ at every $W$. Its inner impostors are still outer folds and no trial was refitted at full budget, so the configuration remains one that works rather than one shown to be optimal --- though no longer for want of an independent search.

\subsection{Moving Average and Decision}\label{sec:postprocessing}
Because a single keystroke carries instantaneous noise arising from finger fluctuation, the system applies a moving average to the probability scores emitted by the softmax layer.

For a sequence of raw probability scores $S = \{s_1, s_2, \dots, s_n\}$ corresponding to consecutive keystrokes, the smoothed score $\bar{s}_t$ at time step $t$ is computed with the smoothing window size $W$ as
\begin{equation}
\bar{s}_t = \frac{1}{W} \sum_{i=0}^{W-1} s_{t-i}, \qquad t \ge W.
\end{equation}

The identity is accepted if $\bar{s}_{t} \ge \tau$, one decision per smoothed score. $\tau$ is swept rather than fixed, since the equal error rate and the false rejection rate at $\mathrm{FAR}=0.1\%$ each define their own (Section~\ref{sec:evaluation_metrics}).

Two boundary conditions are worth stating, since they determine how many decisions the system emits. First, the average is strictly trailing and is formed only once $W$ scores are available, so the first $W-1$ keystrokes of a sequence produce no authentication decision; at $W=3$ this is the cost of the approximately 0.77\,s enrollment latency quantified in Section \ref{sec:latency_model}. Second, the buffer is reset whenever the underlying identity changes, so that no smoothed score is ever formed from a mixture of genuine and impostor keystrokes. The number of smoothed scores available for evaluation therefore decreases as $W$ grows, which is why we do not report windows beyond $W=7$.


\section{Data Collection and Evaluation}
\label{sec:evaluation}
We name the protocol \emph{leave-one-impostor-out} rather than leave-one-subject-out because the enrolled genuine user is fixed and only the impostor role rotates: only the target user was recorded in both domains (Section \ref{sec:limitations}). What follows therefore characterizes one enrolled user against nine unseen attackers, and the reported spread is a spread over attackers rather than over enrolled users.

Two protocols appear below. The \emph{rotated} protocol repeats the evaluation with each of the nine remaining participants as the held-out impostor, and is the only one supporting a claim about unseen impostors in general. The \emph{single-configuration} protocol fixes one held-out impostor and varies only the factor under study. Figures from the two are not interchangeable: the rotated mean at $W=3$ is 0.53\% against 0.25\% for the single configuration.

\subsection{Data Collection Setup}
\label{sec:data_collection}

\subsubsection{Data Collection Environment}
We collected typing data from ten participants, denoted P01--P10, with P01 enrolled as the genuine user and P02--P10 acting as impostors. All were right-handed and of similar hand size, so that the custom glove fitted properly. Each wore the glove while typing on one of two keyboards: a standard membrane keyboard (Lite-On UCL111UBK1) on a desktop machine, or the built-in keyboard of a MacBook. P01 typed on both, on separate days; P10 typed on the MacBook only and the remaining eight on the desktop keyboard only.

\subsubsection{Participant Consent and Data Handling}
All participants were adults, informed before taking part of the purpose of the study, the signals to be recorded, the intended use of the data and their right to withdraw at any point without giving a reason; all consented on that basis. Participation was voluntary, unpaid and noninvasive. No personally identifying information beyond the sensor and keystroke recordings was collected, and the task was restricted to transcribing three prescribed sentences, so no free text of the participants' own composition was captured. Recordings are held under these pseudonymous identifiers and the raw data are not published, for the reasons given in Section \ref{sec:conclusion}.

\subsubsection{Typing Task}
The task comprised three English sentences:

\begin{itemize}
    \item \textbf{Pangram} -- ``Quick brown foxes jump lazily over the slick, icy roads.'' Covers most of the alphabet, so transitions across the layout are varied.
    \item \textbf{Complexity} -- ``We must examine the complex physics of typing rhythms and pressure.'' Demanding key sequences.
    \item \textbf{Natural flow} -- ``The time and motion of every human touch on the keys is unique.'' Unaffected typing.
\end{itemize}

Each participant typed each sentence ten times.

\subsubsection{Dataset Partitioning}
Typing with a data glove requires a brief adaptation period, so the first repetition of the first sentence in each recording session was excluded; all subsequent repetitions were retained. This filter removes 825 of the 22{,}700 recorded key presses, leaving 21{,}875 keystroke events and 21{,}855 feature windows. The Leave-One-Impostor-Out (LOIO) partition is then:

\begin{enumerate}
    \item \textbf{Training set:} P01's desktop session as the positive class, and eight of the nine impostors as the negative class. All are desktop recordings except P10's, which exists only on the laptop.
    \item \textbf{Validation set:} the final 20\% of each training recording, used solely for early stopping.
    \item \textbf{Test set:} P01's laptop session, recorded on a separate day, against the one impostor excluded from training.
\end{enumerate}

Only P01 and P10 were recorded on the laptop keyboard. In every fold except the one that holds P10 out, genuine laptop data is therefore paired with a desktop impostor, and P10's laptop recording sits in the training pool carrying the impostor label. In eight folds of nine, a change of person is also a change of device and day. Section \ref{sec:channel_control} examines which the classifier separates.

\subsection{Evaluation Metrics}\label{sec:evaluation_metrics}
Following the standard definitions in the keystroke dynamics literature \cite{Roy2022}, we quantify performance in terms of the False Acceptance Rate (FAR), the False Rejection Rate (FRR) and the Equal Error Rate (EER), which are defined as follows:

\begin{equation}
    \mathrm{FAR} = \frac{\mathrm{FP}}{\mathrm{TN} + \mathrm{FP}}, \label{eq:far}
\end{equation}

\begin{equation}
    \mathrm{FRR} = \frac{\mathrm{FN}}{\mathrm{TP} + \mathrm{FN}}, \label{eq:frr}
\end{equation}

where $\mathrm{TP}$ and $\mathrm{TN}$ count genuine attempts correctly accepted and impostor attempts correctly rejected, and $\mathrm{FP}$ and $\mathrm{FN}$ their respective errors. Both rates are functions of the decision threshold applied to the smoothed score, and the EER is the operating point at which they coincide. Because the score distribution is discrete the two curves need not intersect exactly, so we take the threshold minimizing their absolute difference and report the false acceptance rate there.

We additionally report the false rejection rate at a fixed, security-oriented operating point, $\mathrm{FAR} = 0.1\%$, which is what a deployment would be tuned to. Since no observed threshold attains that rate exactly, the value is obtained by linear interpolation between the two operating points that bracket it.

\subsection{Decision Latency}
\label{sec:latency_model}
The end-to-end latency of a smoothed decision has two sources: the feature window $T_{w}$ imposes a fixed per-event tail latency, since a feature vector cannot be finalized until $T_{w}/2$ after the key press, and accumulating $W$ consecutive scores requires waiting for those keystrokes to occur, which depends on typing speed. The total is
\begin{equation}
T_{d}(W) = (W-1)\,\Delta t_{k} + \frac{T_{w}}{2}, \label{eq:latency}
\end{equation}
where $\Delta t_{k}$ is the interval between consecutive key presses, determined by human typing pace and unrelated to the 10-ms inter-packet interval of the glove. We adopt its median over the corpus ($232.6$\,ms) rather than the mean ($360.1$\,ms), which is inflated by occasional pauses. Latency computed this way is not a measurement of a running implementation.

\subsection{Experimental Results}
\label{sec:experimental_results}

\subsubsection{Overall Robustness Evaluation (Rotated Protocol)}
Each fold is repeated with three random seeds. Results are aggregated first as the mean over seeds within a fold, then as the mean and standard deviation across folds. Table \ref{tab:overall_results} reports moving-average window sizes $W$ from 1 to 7.

\begin{table}[htbp]
\centering
\caption{Leave-One-Impostor-Out Evaluation Under the Rotated Protocol}
\label{tab:overall_results}
\resizebox{1.0\columnwidth}{!}{
\begin{tabular}{c|cc}
\hline
\textbf{Smoothing} & \textbf{EER (\%)} & \textbf{FRR @ FAR=0.1\% (\%)} \\
\textbf{window} $\mathbf{W}$ & mean $\pm$ SD & mean (worst fold) \\ \hline
1 & 1.55 $\pm$ 1.02 & 13.62 (31.24) \\
2 & 0.88 $\pm$ 0.58 & 6.44 (21.00) \\
\textbf{3} & \textbf{0.53} $\pm$ \textbf{0.42} & 3.09 (11.44) \\
4 & 0.31 $\pm$ 0.31 & 1.71 (9.35) \\
5 & 0.18 $\pm$ 0.22 & 0.68 (4.20) \\
6 & 0.09 $\pm$ 0.15 & 0.49 (3.84) \\
7 & 0.04 $\pm$ 0.08 & 0.16 (1.28) \\ \hline
\end{tabular}
}
\end{table}

Error rates fall monotonically with $W$: the mean EER across folds is 1.55\% without smoothing ($W=1$), 0.53\% at $W=3$ and 0.18\% at $W=5$.

The spread across folds is large, however. At $W=3$ the easiest fold attains 0.09\% and the hardest 1.42\%, a factor of 15 between them. That spread is not training randomness: rerunning a fold with a different seed moves the result by less (0.26\%) than changing the fold does (0.42\%), so what drives the result is which impostor is held out. The hardest is P07 and the easiest P09. This is the biometric menagerie: error concentrates in a minority of subjects rather than spreading over a population \cite{doddington1998sheep, yager2010menagerie}. Our genuine user is fixed, so the folds rank attackers, and an evaluation against one impostor samples a single point of that distribution.

\begin{figure}[htbp]
    \centering
    \includegraphics[width=1.0\linewidth]{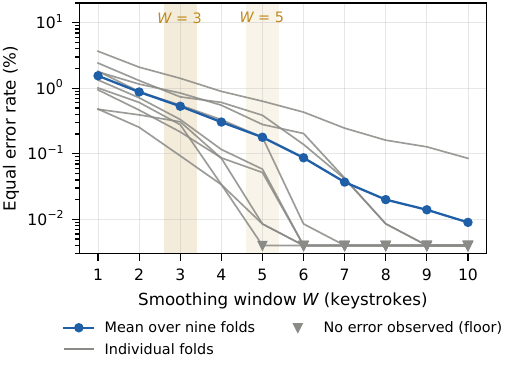}
    \caption{EER against the smoothing window, per held-out impostor.}
    \label{fig:loso_smoothing}
\end{figure}

\begin{figure}[htbp]
    \centering
    \includegraphics[width=1.0\linewidth]{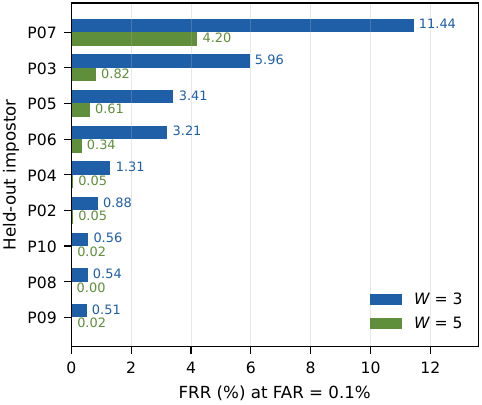}
    \caption{False Rejection Rate at FAR = 0.1\%, per held-out impostor, ordered by $W=3$.}
    \label{fig:loso_frr}
\end{figure}

Figure \ref{fig:loso_smoothing} plots the per-fold EER against the smoothing window, showing how the spread narrows as $W$ grows. Figure \ref{fig:loso_frr} gives the false rejection rate per impostor at the strict operating point, showing how much the difficulty depends on who the attacker is.

The resolution of an estimate based on the $n\approx3{,}950$ smoothed scores of a fold (3{,}857 to 4{,}072 across the nine) is $1/n\approx0.025\%$. One fold observes no error at $W=5$, five do at $W=6$ and six at $W=7$, so the means at those windows increasingly measure the size of the test set rather than the system; no fold is at the floor at $W=3$. We therefore report and recommend $W=3$.
\begin{table*}[b]
\centering
\caption{Architecture Comparison Under the Rotated Protocol (Mean $\pm$ SD Across Folds [\%]), ordered by $W=3$.}
\label{tab:model_comparison}
\resizebox{0.85\textwidth}{!}{
\begin{tabular}{l|cc|ccc}
\hline
\textbf{Model} & $\mathbf{T_{w}}$ & $\mathbf{T_{d}}$ & $\mathbf{W=1}$ & $\mathbf{W=3}$ & $\mathbf{W=5}$ \\
 & (ms) & at $W=3$ (s) & & & \\ \hline
\textbf{CNN+LSTM, two branches (proposed)} & 600 & 0.77 & \textbf{1.55} ($\pm$ 1.02) & \textbf{0.53} ($\pm$ 0.42) & \textbf{0.18} ($\pm$ 0.22) \\
CNN Only & 1200 & 1.07 & 1.04 ($\pm$ 0.90) & 0.59 ($\pm$ 0.66) & 0.33 ($\pm$ 0.46) \\
CNN+LSTM, single stream & 600 & 0.77 & 1.59 ($\pm$ 1.00) & 0.62 ($\pm$ 0.46) & 0.24 ($\pm$ 0.28) \\
Transformer & 1800 & 1.37 & 1.80 ($\pm$ 1.25) & 1.01 ($\pm$ 0.86) & 0.76 ($\pm$ 0.78) \\
CNN+GRU & 1600 & 1.27 & 1.82 ($\pm$ 1.67) & 1.34 ($\pm$ 1.39) & 1.06 ($\pm$ 1.18) \\
TCN & 400 & 0.67 & 3.41 ($\pm$ 2.05) & 1.38 ($\pm$ 1.06) & 0.48 ($\pm$ 0.52) \\
Adapted BehaveFormer STDAT & 800 & 0.87 & 4.15 ($\pm$ 2.46) & 2.98 ($\pm$ 2.04) & 2.25 ($\pm$ 1.61) \\
LSTM Only & 2000 & 1.47 & 3.31 ($\pm$ 3.11) & 3.02 ($\pm$ 2.99) & 2.83 ($\pm$ 2.91) \\ \hline
\end{tabular}
}
\end{table*}

\subsubsection{Visualization of Classification Performance}

\begin{figure}[htbp]
    \centering
    \includegraphics[width=0.9\linewidth]{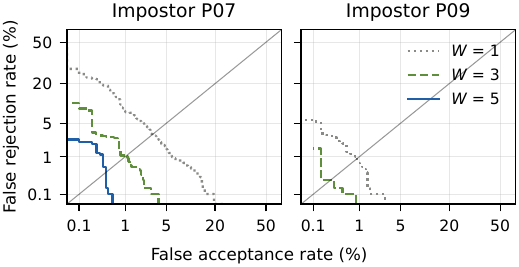}
    \caption{DET curves for the hardest and the easiest held-out impostor.}
    \label{fig:det_curve_comp}
\end{figure}

A Detection Error Tradeoff (DET) curve is defined for a single score set and cannot be averaged over folds, so Figure \ref{fig:det_curve_comp} shows two individual folds of the rotated protocol, the hardest and the easiest held-out impostor, comparing $W=1, 3$ and $5$ on a logarithmic scale. It illustrates the shape of the trade-off; the magnitudes are those of Table \ref{tab:overall_results}.

\subsection{Comparative Analysis and Ablation Studies}
\label{sec:comparative_ablation}

We compared the proposed CNN+LSTM model against ablation baselines (CNN Only and LSTM Only), representative temporal models (CNN+GRU, TCN, and a generic Transformer), and the task-specific BehaveFormer spatio-temporal dual-attention Transformer \cite{Senarath2023}.

For BehaveFormer we adapted the Spatio-Temporal Dual Attention Transformer (STDAT) from the authors' reference implementation (\url{https://github.com/DilshanSenarath/BehaveFormer}). As in the published design, two branches encode the two modalities, here the twenty pressure-side channels and the nine IMU channels; we attach a binary classification head and, with it, replace the triplet-loss objective around which STDAT was designed by the class-weighted cross-entropy of \eqref{eq:loss}. Because that changes the training objective and not merely the output layer, we call it an adapted STDAT variant (Section \ref{sec:sota_comparison}).

Table \ref{tab:model_comparison} reports the mean EER of each model at $W=1, 3$ and $5$.

Every row uses the same 29 channels, and each architecture keeps the window and the interpolated length its own search selected ($T_{w}$ from 400 to 2000\,ms, $L$ from 30 to 45). The baselines carry a standardized pressure-level block and the proposed model does not, because standardizing it changes nothing for two branches (Section \ref{sec:modality_ablation}).

The proposed model is not significantly better than CNN Only, the single stream (Wilcoxon signed-rank over the folds, $p=0.50$ in both cases at $W=3$), the Transformer or CNN+GRU ($p=0.074$ in both). It is significantly better than the TCN ($p=0.0078$), LSTM Only ($p=0.0195$) and the adapted BehaveFormer STDAT ($p=0.0039$).

\begin{figure*}[t]
    \centering
    \includegraphics[width=\textwidth]{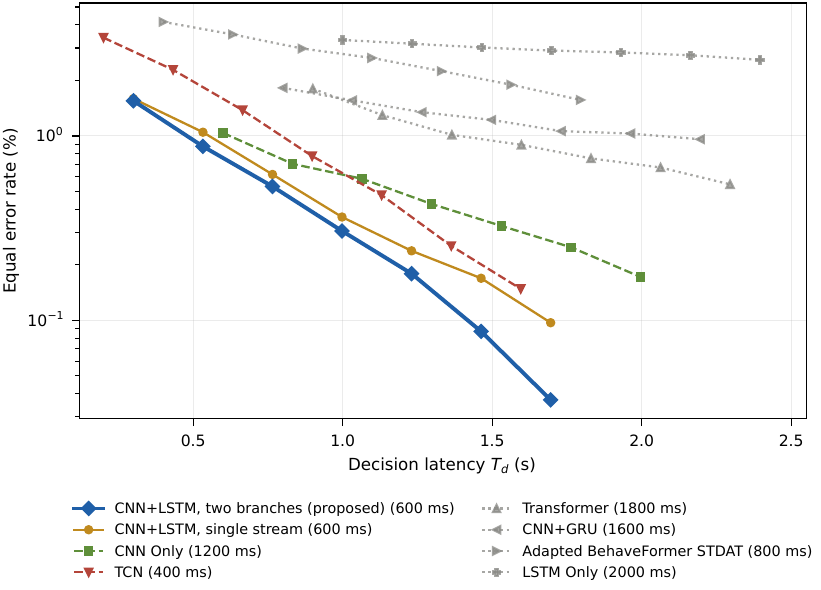}
    \caption{Equal error rate against decision latency, rotated protocol, $W=1$ to $7$.}
    \label{fig:latency_pareto}
\end{figure*}

CNN Only needs a 1200\,ms window, so its 0.59\% arrives at 1.07\,s, while the proposed model reaches 0.53\% at 0.77\,s and 0.31\% at 1.00\,s. The single stream shares the 600\,ms window and so the same latency. The TCN alone decides faster, always at more than twice the error rate. Figure \ref{fig:latency_pareto} shows the trade-off for all eight. LSTM Only is given the longest window of all at 2000\,ms and no convolutional front end, and is still the weakest at 3.02\%: a wider window does not substitute for the cross-channel structure a convolution extracts. The proposed model reaches the lowest error at the shortest window among the accurate models.

\subsection{Feature-Window Length}
\label{sec:window_analysis}

\subsubsection{Window Overlap}
Consecutive keystrokes fall closer together than 600\,ms in $83.6\%$ of cases, so adjacent windows share the same span of sensor samples -- on average $50.2\%$ of a window. No information reaches the test set this way, because windows are built inside a single recording unit and the test sessions contribute nothing to training or to model selection. The one boundary that overlapping windows do cross is the train/validation split taken chronologically inside each training unit, where a handful of windows per unit share samples; that partition is used for early stopping alone and never for reporting. Per-keystroke scores are, however, correlated, so the gain from smoothing cannot be derived analytically, which is why the error rates of Table \ref{tab:overall_results} are measured at every $W$.

\subsubsection{Sensitivity to the Window Length}
We retrained the proposed model for ten settings of $T_{w}$ between 400 and 1000\,ms, under the rotated protocol with three seeds in each of the nine folds. The single-configuration protocol cannot serve here: against its held-out impostor, P10, the model scores at or near zero for every window from 500\,ms upward, so the settings cannot be separated.

\begin{table}[htbp]
\centering
\caption{Sensitivity of EER (\%) to the Feature Window Length $T_{w}$}
\label{tab:window_sensitivity}
\resizebox{1.0\columnwidth}{!}{
\begin{tabular}{c|ccc}
\hline
$T_{w}$ (ms) & $W=1$ & $W=3$ & $W=5$ \\ \hline
400 & 11.44 $\pm$ 7.88 & 7.69 $\pm$ 7.32 & 5.53 $\pm$ 6.04 \\
450 & 7.13 $\pm$ 5.69 & 4.21 $\pm$ 4.73 & 2.95 $\pm$ 3.90 \\
500 & 2.52 $\pm$ 1.87 & 0.98 $\pm$ 0.93 & 0.34 $\pm$ 0.49 \\
550 & 2.00 $\pm$ 1.06 & 0.68 $\pm$ 0.58 & 0.24 $\pm$ 0.34 \\
\textbf{600 (used)} & 1.55 $\pm$ 1.02 & 0.53 $\pm$ 0.42 & 0.18 $\pm$ 0.22 \\
650 & 1.05 $\pm$ 0.54 & 0.39 $\pm$ 0.32 & 0.13 $\pm$ 0.17 \\
700 & 1.04 $\pm$ 0.70 & 0.42 $\pm$ 0.34 & 0.16 $\pm$ 0.20 \\
800 & 1.01 $\pm$ 0.73 & 0.41 $\pm$ 0.34 & 0.16 $\pm$ 0.19 \\
900 & 1.20 $\pm$ 0.97 & 0.47 $\pm$ 0.52 & 0.22 $\pm$ 0.32 \\
1000 & 1.10 $\pm$ 1.17 & 0.51 $\pm$ 0.67 & 0.26 $\pm$ 0.46 \\ \hline
\end{tabular}
}
\end{table}

The $T_{w}=600$\,ms row is the configuration of Table \ref{tab:overall_results} in every respect and reproduces it at 0.53\% and 0.18\%, serving as a reference arm.

\begin{figure}[htbp]
\centering
\includegraphics[width=1.0\linewidth]{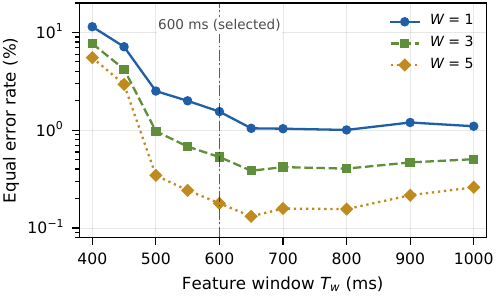}
\caption{Sensitivity of EER to the feature-window length (rotated protocol).}
\label{fig:window_sensitivity}
\end{figure}

The result is a wide basin. At $W=3$ and $W=5$ only 400 and 450\,ms are significantly worse than 600\,ms (Wilcoxon signed-rank with Holm correction, $p=0.035$ in both cases and worse in every fold), and everything from 500 to 1000\,ms is indistinguishable from it. At $W=1$ the floor sits higher: 500 and 550\,ms are significantly worse as well ($p=0.035$, worse in every fold), and only 650\,ms upward is indistinguishable.

The numerical minimum falls at 650\,ms, 0.39\% against 0.53\% at $W=3$, but that difference is not significant ($p=1.0$ after correction). Stretching the window to 1000\,ms adds only 0.2\,s to a decision. $T_{w}=600$\,ms is therefore not a sharp optimum but a point inside a range over which both accuracy and latency are flat; what the sweep does establish is a floor at 500\,ms, below which accuracy falls steeply.

The floor is where the neighboring keystrokes enter the window: centered on the key press, the window admits them once its half-width exceeds the median inter-keystroke interval of 232.6\,ms, and the measured cliff falls at the $T_{w} \approx 465$\,ms this predicts. Window-length studies in wearable sensing report the same shape and the same rule, to take the shortest window inside the saturated range \cite{banos2014window}.

\subsection{Modality and Fusion Ablation}
\label{sec:modality_ablation}
We repeated the rotated evaluation over each modality alone and over both, encoded as a single stream and as two branches, with and without the ten pressure-level channels. The folds, splits, hyperparameters, smoothing and metrics are those of Table \ref{tab:overall_results}; only the input and the encoder vary, and the folds are built once and sliced along the channel axis so that every arm sees the same windows. Because Z-score normalization is computed per window and per channel, a channel's normalized values do not depend on which others are present, so slicing after normalization is exact. Tables \ref{tab:modality_ablation} and \ref{tab:level_encoder} report the result.

\begin{table}[htbp]
\centering
\caption{Modality Ablation Under the Rotated Protocol (Mean $\pm$ SD Across Folds [\%]).}
\label{tab:modality_ablation}
\resizebox{1.0\columnwidth}{!}{
\begin{tabular}{l|ccc}
\hline
\textbf{Input channels} & $\mathbf{W=1}$ & $\mathbf{W=3}$ & $\mathbf{W=5}$ \\ \hline
Pressure only (20 ch) & 6.82 ($\pm$ 6.62) & 4.44 ($\pm$ 4.55) & 3.16 ($\pm$ 3.59) \\
IMU only (9 ch) & 4.49 ($\pm$ 3.25) & 2.28 ($\pm$ 2.31) & 1.14 ($\pm$ 1.60) \\
\textbf{Both, two branches} & \textbf{1.55} ($\pm$ 1.02) & \textbf{0.53} ($\pm$ 0.42) & \textbf{0.18} ($\pm$ 0.22) \\ \hline
\end{tabular}
}
\end{table}

\begin{table}[htbp]
\centering
\caption{Pressure Level and Encoder Under the Rotated Protocol (Mean $\pm$ SD Across Folds [\%]).}
\label{tab:level_encoder}
\resizebox{1.0\columnwidth}{!}{
\begin{tabular}{ll|ccc}
\hline
\textbf{Pressure level} & \textbf{Encoder} & $\mathbf{W=1}$ & $\mathbf{W=3}$ & $\mathbf{W=5}$ \\ \hline
absent & single stream & 3.83 ($\pm$ 2.00) & 1.75 ($\pm$ 1.28) & 0.70 ($\pm$ 0.69) \\
absent & two branches & 3.70 ($\pm$ 2.44) & 1.74 ($\pm$ 1.59) & 0.73 ($\pm$ 0.92) \\
raw & single stream & 6.53 ($\pm$ 6.22) & 4.20 ($\pm$ 4.19) & 2.79 ($\pm$ 2.97) \\
raw & \textbf{two branches} & \textbf{1.55} ($\pm$ 1.02) & \textbf{0.53} ($\pm$ 0.42) & \textbf{0.18} ($\pm$ 0.22) \\
standardized & single stream & 1.71 ($\pm$ 1.08) & 0.69 ($\pm$ 0.57) & 0.26 ($\pm$ 0.29) \\
standardized & two branches & 1.72 ($\pm$ 1.13) & 0.63 ($\pm$ 0.49) & 0.21 ($\pm$ 0.24) \\ \hline
\end{tabular}
}
\end{table}

Table \ref{tab:modality_ablation} answers what each modality contributes. The fused model reaches 0.53\% EER at $W=3$ against 2.28\% for the inertial channels alone and 4.44\% for the pressure channels alone. Compared fold by fold with a Wilcoxon signed-rank test over the nine held-out impostors, the fused model beats the inertial branch in all nine folds ($p=0.0039$) and the pressure branch in seven ($p=0.027$); at $W=5$ the corresponding figures are eight of nine ($p=0.0078$) and eight of nine ($p=0.020$). The advantage holds at every $W$ up to 7 ($p$ at most 0.039 against the inertial branch and 0.027 against the pressure branch). Neither modality is redundant. The two are complementary in the strict sense that each contributes something the other does not.

With the level block raw, a single stream reaches 4.20\% at $W=3$, worse than the inertial channels alone, while two branches on the same input reach 0.53\% (Table \ref{tab:level_encoder}). Standardizing the block on training statistics brings the single stream to 0.69\%, better in every fold ($p=0.0039$). Two branches achieve the same structurally, and hold with or without the transform ($p=0.30$). Scale-matched, the two architectures are indistinguishable ($p=0.84$).

\begin{figure}[htbp]
    \centering
    \includegraphics[width=1.0\linewidth]{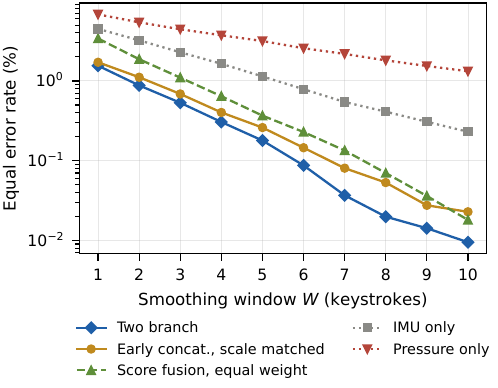}
    \caption{Three fusion designs and the two single-modality models, under the rotated protocol.}
    \label{fig:fusion_designs}
\end{figure}

The cause is the competition for the shared first layer described in Section \ref{sec:model}. The single stream shows it in opposite directions with the level block raw and without it; Figure \ref{fig:fusion_designs} places the standardized single stream beside the two other fusion designs we tried. With the pressure-level channels present, the fused per-fold errors track the pressure-only arm at $r=0.977$ and the inertial one at $r=-0.184$ ($W=5$); with the nineteen Z-scored channels the same correlations are $-0.085$ and $0.976$, so there it is the inertial channels that are followed and the pressure channels ignored. Either way the fused model reproduces one arm and not the other, which is what capture rather than fusion looks like.

Multimodal networks are known to be dominated by whichever stream is easier to fit, the joint model tracking that stream while the other contributes little \cite{wang2020multimodal, peng2022balanced}. What this ablation adds is that magnitude alone induces it: the level block carries the same information standardized or not, and standardizing it removes the capture (Table \ref{tab:level_encoder}). A per-modality ablation is therefore necessary to establish complementarity but not sufficient to refute it: a null result from one architecture is a statement about that architecture, and about its preprocessing.

\subsection{Channel Controls: Identity or Recording Session?}
\label{sec:channel_control}
Two groups could encode the recording rather than the hand: the magnetometer, and the ten pressure-level channels, the only input per-window Z-scoring does not neutralize. We removed each and re-ran the rotated evaluation.

\begin{table}[htbp]
\centering
\caption{Channel Controls Under the Rotated Protocol (Mean $\pm$ SD Across Folds [\%]).}
\label{tab:channel_control}
\resizebox{1.0\columnwidth}{!}{
\begin{tabular}{cc|c|ccc}
\hline
\textbf{Pressure level} & \textbf{Magnetometer} & \textbf{ch} & $\mathbf{W=1}$ & $\mathbf{W=3}$ & $\mathbf{W=5}$ \\ \hline
\checkmark & \checkmark & 29 & 1.55 ($\pm$ 1.02) & 0.53 ($\pm$ 0.42) & 0.18 ($\pm$ 0.22) \\
\checkmark & --- & \textbf{26} & \textbf{1.12} ($\pm$ 0.97) & \textbf{0.37} ($\pm$ 0.50) & \textbf{0.15} ($\pm$ 0.27) \\
--- & \checkmark & 19 & 3.70 ($\pm$ 2.44) & 1.74 ($\pm$ 1.59) & 0.73 ($\pm$ 0.92) \\
--- & --- & 16 & 3.01 ($\pm$ 2.62) & 1.33 ($\pm$ 1.63) & 0.51 ($\pm$ 0.86) \\ \hline
\end{tabular}
}
\end{table}

The magnetometer has the largest raw between-recording shift of any channel ($+199$, $-90$, $-122$~$\mu$T). Removing it improves the model at every $W$, so it is not carrying the result.

The pressure level is worth a factor of 3.3 ($p=0.0078$). The recording does not explain it: the device-matched fold gains on the same scale as the rest (3.0 against a median of 3.9, within the 0.8 to 6.8 range the nine folds span).

\section{Discussion}
\label{sec:discussion}

\subsection{Comparison with State-of-the-Art Methods of Similar Characteristics}
\label{sec:sota_comparison}
Two kinds of comparison are available here and they differ in weight. Our strictest benchmark is the same-protocol comparison of Table \ref{tab:model_comparison}, in which the proposed method outperforms the adapted BehaveFormer baseline in every fold ($p=0.0039$ at $W=3$). That margin confounds architecture with objective and this experiment cannot separate the two, because the adapted baseline departs from the published design in the output layer and in the training objective alike (Section \ref{sec:comparative_ablation}). It is therefore evidence that an STDAT trained as a binary classifier on 29 glove channels from a single enrolled user does not match the proposed hybrid; it is not evidence about BehaveFormer as its authors specify it. Dual-attention Transformers plausibly also require larger training sets than this single-subject setting provides.

We deliberately refrain from juxtaposing our EER with the figures reported for HMOG \cite{sitova2016hmog} and BehaveFormer \cite{Senarath2023}. Both were measured on smartphone touchscreen typing, which differs from our physical keyboard and glove setup in input modality, sensor placement, and threat model; a direct numerical comparison would therefore be misleading.

For a closer match in modality, the self-powered intelligent keyboard of Zhang et al.\ \cite{Zhang2025} instruments a physical keyboard with a flexible pressure-sensor array and reports a 95.3\% success rate on fixed eight-character passwords and 100\% accuracy on dynamic text. This comparison necessarily remains qualitative, since accuracy and EER are not interchangeable metrics. For two further instrumented-keyboard approaches \cite{chen2015personalized, wei2022all}, no independently verifiable error rates were available, and we therefore report none. Where a numerical comparison is not available, the behaviors we observe are placed against the literature instead, in Sections \ref{sec:experimental_results}, \ref{sec:window_analysis} and \ref{sec:modality_ablation}.

\subsection{Security, Usability, and Latency Trade-off}
\label{sec:latency_tradeoff}
Under a strict security setting (FAR = 0.1\%), the mean False Rejection Rate is 3.09\% at $W=3$ and 0.68\% at $W=5$. The spread across held-out impostors is wider than the EER figures alone suggest -- at $W=3$ the best fold attains 0.51\% and the worst 11.44\% -- so a deployment tuned to this point would still reject roughly one genuine attempt in nine against the hardest impostor in our pool, and 4.20\% at $W=5$. For a deployment that must guarantee usability against any impostor the threshold should therefore be calibrated per user against the observed genuine score distribution rather than set globally, with the worst-case fold reported alongside the mean. Larger smoothing windows drive the error rates toward the resolution floor of the test set, but the evidence they accumulate has to be waited for, and that wait is authentication delay.

At $T_{w}=600$\,ms, \eqref{eq:latency} yields $0.77$\,s ($W=3$), $1.0$\,s ($W=4$) and $1.23$\,s ($W=5$); the corresponding values at the mean pace, $1.0$, $1.4$ and $1.7$\,s, serve as an upper bound.

We recommend $W=3$: 0.53\% EER after 0.77\,s, and quote $W=5$ alongside it because a deployment willing to wait does reach 0.18\% after 1.23\,s.

\subsection{Limitations}
\label{sec:limitations}
First, the dataset comprises ten participants, all right-handed and selected for similar hand size in order to fit the custom glove, which limits the generalizability of the absolute EER values to a broader population. The difficulty of the unseen-impostor condition also varies substantially with the identity of the held-out individual (Figures \ref{fig:loso_smoothing} and \ref{fig:loso_frr}): at $W=3$ the easiest and hardest folds differ by a factor of 15, so a single-impostor estimate may fall anywhere between them.

Second, the cross-domain claim rests on one enrolled user in two sessions, and only one fold compares classes recorded on the same device. A linear probe trained to separate P01 from the impostors on desktop data alone scores his laptop windows 0.58 impostor standard deviations further into the genuine region on the pressure-level channels, so part of the cross-domain margin is the recording rather than the person. Section \ref{sec:channel_control} bounds this; a cohort recorded in both domains would remove it.

Third, the evaluation uses the fixed-text transcription of three prescribed sentences; free-text typing is a more demanding condition to which our results do not extend without further validation.

Fourth, the method requires the user to wear an instrumented glove, which is a substantial barrier to everyday use: the effort of putting it on, its comfort over long sessions and the durability of the custom hardware are all unevaluated. Miniaturization remains future work (Section \ref{sec:introduction}).

\section{Conclusion}
\label{sec:conclusion}

We have proposed a timing-free multimodal authentication system in which 19 glove-borne pressure and inertial channels are encoded by two CNN-LSTM branches and smoothed over consecutive keystrokes. Under a cross-domain Leave-One-Impostor-Out protocol, three keystrokes -- about 0.77 seconds of typing -- yield a mean Equal Error Rate of 0.53 $\pm$ 0.42\% and a mean False Rejection Rate of 3.09\% at a False Acceptance Rate of 0.1\%, falling to 0.18 $\pm$ 0.22\% and 0.68\% at $W=5$.

The spread across impostors remains wide, so a deployment must calibrate the threshold per user. A per-modality ablation showed each modality to carry identity information the other does not supply, and showed the pressure level to be worth a factor of 3.3 provided it reaches the classifier on a comparable scale. A block of channels much larger than the rest dominates whatever layer mixes them, and matching its scale is worth as much as separating the encoders. Future work will collect a laptop session for every participant, expand the cohort in scale and demographic breadth, extend the evaluation to free-text entry, search the fusion mechanism itself, and optimize the model for real-time authentication on wearable edge devices.

Our implementation -- preprocessing, the proposed model, the training and evaluation scripts, and all baselines -- is publicly available at \url{https://github.com/Issei-hyakuda/multimodal-keystroke-glove-authentication}.
The raw recordings are withheld under the informed-consent protocol governing their collection, and are available from the corresponding author upon reasonable request, subject to participant consent.

\bibliographystyle{IEEEtran}
\bibliography{references}


\begin{IEEEbiography}[{\includegraphics[width=1in,height=1.25in,clip,keepaspectratio]{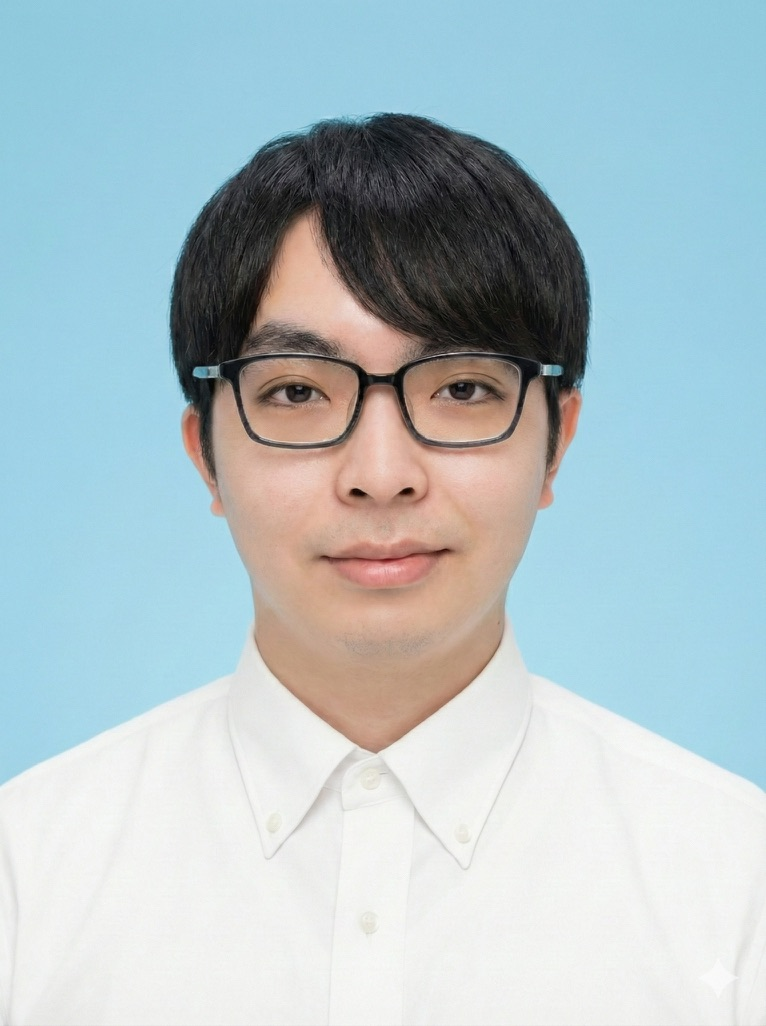}}]{Issei Hyakuda}
received the B.S. degree from the University of Aizu, Aizu-Wakamatsu, Japan, in 2026. He is currently working
toward the M.S. degree with the Graduate School
of Computer Science and Engineering, University of Aizu, Aizu-Wakamatsu, Japan.
His research interests include biometric authentication and human-computer interaction.
\end{IEEEbiography}

\begin{IEEEbiography}[{\includegraphics[width=1in,height=1.25in,clip,keepaspectratio]{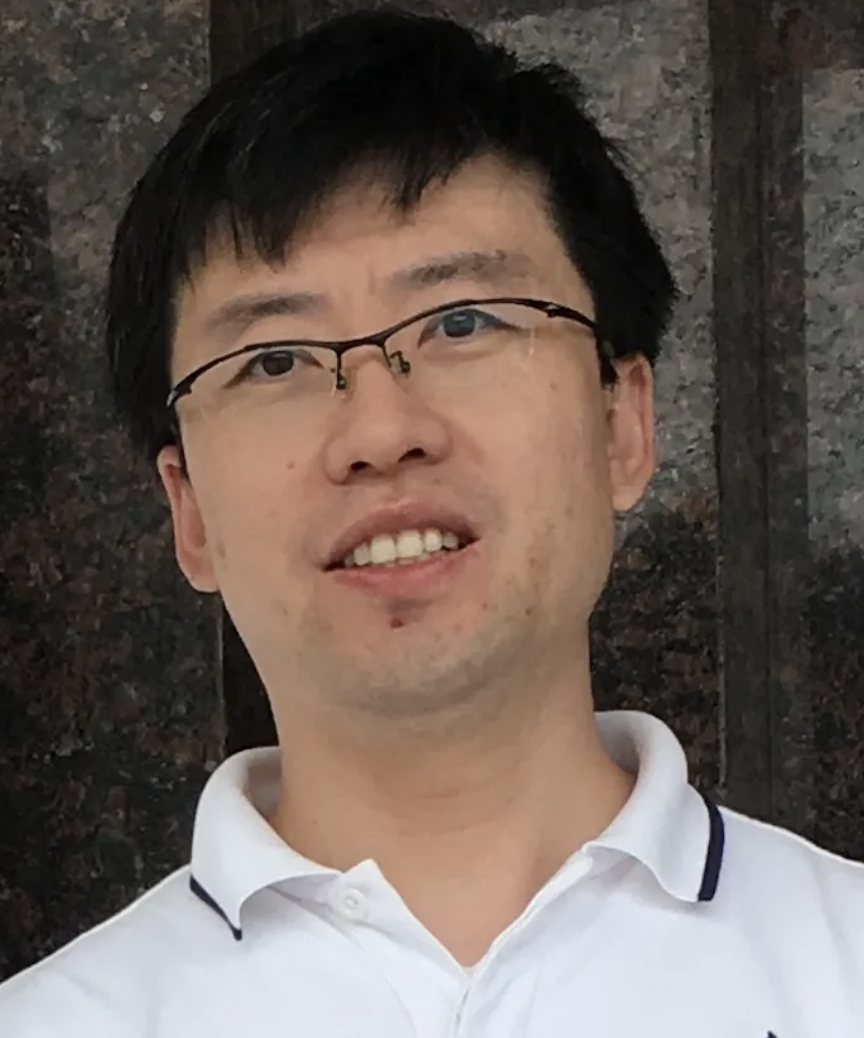}}]{Lei Jing}
(M'12) received the Ph.D. degree in
computer science and engineering from the University of Aizu, Aizu-Wakamatsu, Japan, in 2008.
He is currently a Senior Associate Professor with the School of
Computer Science and Engineering, University
of Aizu. His research interests include human
position, posture, and motion tracking, soft circuit design, and the tactile internet. Applications of his work encompass the detection of abnormalities in human activity, sign language recognition,
and human-robot interaction. He has published
over 120 papers and holds six patents in related areas.
\end{IEEEbiography}

\end{document}